\documentclass[preprint, double-spaced]{revtex4}
\usepackage{bbm}
\usepackage{mathrsfs}
\usepackage{epsfig,psfrag}
\usepackage[final]{pdfpages}
\usepackage{amsmath,amsfonts,amssymb}
\usepackage{graphicx}
\usepackage[dvips, bookmarks, colorlinks=true, plainpages = false, citecolor = blue, linkcolor = blue, urlcolor = blue, filecolor = blue]{hyperref}
\def\be{\begin{equation}}
\def\ee{\end{equation}}
\def\bea{\begin{eqnarray}}
\def\eea{\end{eqnarray}}

\newcommand{\us}{\uparrow}
\newcommand{\ds}{\downarrow}

\begin{document}


\title{Antikink dispersions of the $J_1$-$J_2$ sawtooth spin-1/2 anisotropic Heisenberg antiferromagnetic chain}
\author{Susobhan Paul}\email{suso.phy.paul@gmail.com}
\affiliation {Department of Physics, Scottish Church College, 
Urquhart Square, Kolkata 700006, India}
\author{Asim Kumar Ghosh}
 \email{asimkumar96@yahoo.com}
\affiliation {Department of Physics, Jadavpur University, 
188 Raja Subodh Chandra Mallik Road, Kolkata 700032, India}

\date{\today}

\begin{abstract}
Antikink dispersion of the asymmetric $J_1$-$J_2$ sawtooth spin-1/2 
anisotropic Heisenberg antiferromagnetic chain has been derived. 
Value of spin-gap is obtained by estimating the minimum of dispersion relation 
variationally. The exact doubly-degenerate ground state 
energy has been 
derived at the symmetric point, $J_1=J_2$, for the whole anisotropic regime. 
Analytic form of dispersion relations of three different antikink states is  
obtained and their validity in the parameter space is discussed. 
The value of spin-gap, specific heat and 
susceptibility are estimated numerically by diagonalizing the 
Hamiltonian for chains of finite length. 
The inherent frustration of this antiferromagnetic model leads to 
the appearance of 
additional peak at low-temperatures in the specific heat.  
\end{abstract}
\maketitle
\section{Introduction}
Investigation on the properties of topological excitations for the 
Heisenberg antiferromagnetic (AFM) systems is going on through the 
last several decades. The spin-1/2 Heisenberg AFM chain exhibits 
spinon excitations with no spin-gap \cite{Pearson,Takhtajan}, 
while the Ising-like anisotropic chain 
demonstrates AFM domain-wall excitations with non-zero spin-gap 
\cite{McCoy,Villain,Ghosh1}. Spinons are basically 
spin-1/2 modes and they always appear in pairs. 
Domain-wall states are formed by joining two oppositely oriented 
AFM domains. In 1981, Shastry and Sutherland have introduced a new class of 
topological spin excitations those contain isolated defects between two different 
regions of broken translational symmetry \cite{Shastry-Sutherland}. 
Those spin excitations are found to constitute the low-energy modes of 
a frustrated spin chain which is now known as the Majumder-Ghosh (MG) model \cite{Majumdar-Ghosh}. 
Two different kinds of topological modes called kink and antikink states are 
constructed whose dispersions are similar and both produce the same 
value of spin-gap for the MG model \cite{Caspers}, though they are not the 
eigenstates. 
However, in case of sawtooth \cite{Sen-Shastry} (or $\Delta$ \cite{Nakamura}) chain 
low-energy kink and antikink states possess dissimilar properties. 
Here, kinks are dispersionless as well as gapless, while antikinks 
are dispersive and yield a spin-gap. The Heisenberg interactions in the sawtooth chain 
can be visualized as a chain of triangles with no vertex-vertex interaction as shown in 
Fig.\ref{sawtooth} (a), but with the same bond strength. 
Transition from sawtooth to MG model can be made possible  
by invoking the vertex-vertex interactions and tuning the bond strengths accordingly. 
Crossover from sawtooth to MG model has been studied before \cite{Chen}. 

In the delafossite compounds, RCuO$_{\rm 2+x}$ (R=Y, La, etc), 
spin-1/2 Cu$^+$ ions form an array of planar triangles \cite{Cava}. Thus,  
it has been considered as one of the physical realization of the sawtooth chain. 
However, on doping, the additional x = 0.5 O ions, located at the centers of the alternate 
triangles may alter the values of exchange strengths for base and the 
arm bonds leading to the realization of asymmetric sawtooth chain.  
Thus, in the asymmetric sawtooth chain, the exchange strength for 
the base of the triangle is different from those for the 
other two arms of that. Another realization of magnetic sawtooth lattice is 
the compounds olivines, ZnL$_2$S$_4$ (L=Er,Tm,Yb) \cite{Lau}.
In 2004, Blundell and N\'u\~nez-Regueiro numerically studied  
the asymmetric sawtooth chain and showed that the value of spin-gap varies with 
the difference of base-arm bond strengths \cite{Blundell}. Parameter regime with non-zero 
spin-gap has been identified for the asymmetric sawtooth chain 
at the isotropic point of AFM Heisenberg interactions. Enhancement of spin-gap by 
relaxing the frustration in terms of distorted bond strength 
of sawtooth lattice model has been reported \cite{Nakamura-Takada}.
In another study, properties of localized magnons of the AFM Heisenberg sawtooth model  
due to the presence of frustration in the system is investigated \cite{Honecker}. 
Other significant observation in the sawtooth chain is the 
existence of two peaks in the specific heat \cite{Nakamura-Takada}. 
Specific heat with double peaks is experimentally found in 
$^3$He adsorbed on a graphite substrate \cite{Busch} which is 
eventually reproduced in the numerical studies on frustrated AFM Heisenberg model for the 
twelve-spin cluster of the kagom\'e lattice \cite{Elser}. So, the 
appearance of additional peak in the specific heat at low temperatures
attributes to the presence of frustration in the AFM Heisenberg sawtooth model. 

In this work, we have considered anisotropic AFM Heisenberg interaction 
on the sawtooth lattice with asymmetric bond strengths and obtained the analytic expressions 
of the antikink dispersions. In order to obtain the value of spin-gap, 
antikinks of three different lengths of cluster 
are considered. The minimum-energy antikink dispersion 
is obtained by following the variational method adopted in \cite{Sen-Shastry}. 
The value of spin-gap is also estimated numerically and compared with the 
analytic results. The effect of frustration in this model is noticed because of the 
emergence of double peaks in the specific heat which is obtained numerically 
for a finite chain. 
In section \ref{S-M}, anisotropic Heisenberg $J_1$-$J_2$ sawtooth model 
is described and the known results are presented. 
Section \ref{A-D} contains the derivations of excitations of three different 
antikink clusters and the spin-gap. 
Results of numerical investigations including the estimation of spin-gap, 
specific heat and magnetic susceptibility 
are presented in section \ref{E-D}. Section \ref{conclusion} 
contains a discussion of the results obtained.
\section{The $J_1$-$J_2$ sawtooth spin-$\frac{1}{2}$ anisotropic Heisenberg chain} 
\label{S-M}
\begin{figure}[h]
\begin{center}
\psfrag{j1}{$J_1$}
\psfrag{j2}{$J_2$}
\psfrag{a}{\text{\scriptsize{(a)}}}\psfrag{b}{\text{\scriptsize{(b)}}}
\psfrag{c}{\text{\scriptsize{(c)}}}\psfrag{d}{\text{\scriptsize{(d)}}}
\psfrag{e}{\text{\scriptsize{(e)}}}\psfrag{f}{\text{\scriptsize{(f)}}}
\psfrag{g}{\text{\scriptsize{(g)}}}\psfrag{h}{\text{\scriptsize{(h)}}}
\psfrag{i}{\text{\scriptsize{(i)}}}
\psfrag{1}{\text{\scriptsize{1}}}\psfrag{2}{\text{\scriptsize{2}}}
\psfrag{3}{\text{\scriptsize{3}}}\psfrag{4}{\text{\scriptsize{4}}}
\psfrag{5}{\text{\scriptsize{5}}}\psfrag{6}{\text{\scriptsize{6}}}
\psfrag{7}{\text{\scriptsize{7}}}\psfrag{8}{\text{\scriptsize{8}}}
\psfrag{9}{\text{\scriptsize{9}}}\psfrag{10}{\text{\scriptsize{10}}}
\psfrag{11}{\text{\scriptsize{11}}}\psfrag{12}{\text{\scriptsize{12}}}
\psfrag{13}{\text{\scriptsize{13}}}\psfrag{14}{\text{\scriptsize{14}}}
\psfrag{15}{\text{\scriptsize{15}}}\psfrag{16}{\text{\scriptsize{16}}}
\psfrag{17}{\text{\scriptsize{17}}}\psfrag{18}{\text{\scriptsize{18}}}
\psfrag{19}{\text{\scriptsize{19}}}
\psfrag{2n}{\text{\scriptsize{$2n$}}}
\psfrag{a1}{\text{\scriptsize{$|2n\rangle _1$}}}
\psfrag{a2}{\text{\scriptsize{$|2n\rangle _2$}}}
\psfrag{a3}{\text{\scriptsize{$|2n\rangle _3$}}}
\psfrag{a4}{\text{\scriptsize{$|2n\rangle _4$}}}
\psfrag{g1}{\text{\scriptsize{$|G_1\rangle$}}}
\psfrag{g2}{\text{\scriptsize{$|G_2\rangle$}}}
\psfrag{k}{\text{\scriptsize{$|K\rangle$}}}
\psfrag{S}{\text{\scriptsize{$|O\rangle=$}}}
\psfrag{a2}{\text{\scriptsize{$|2n\rangle _2$}}}
\psfrag{s}{\text{\scriptsize{$\equiv\frac{1}{\sqrt 2}(\us\ds-\ds\us)$}}}
\includegraphics[scale=0.50]{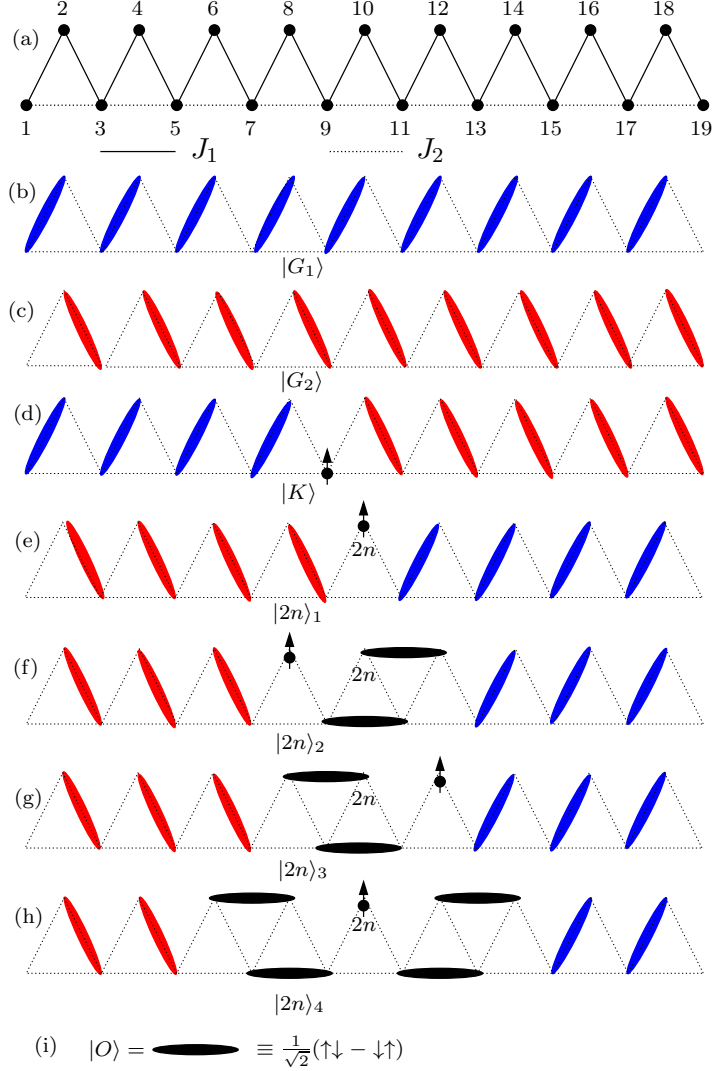}
\caption{\label{sawtooth}{(Color Online) (a) Sawtooth lattice, (b) and (c) ground states $|G_1\rangle$ (blue) and 
$|G_2\rangle$ (red), (d) kink state $|K\rangle$, (e) 1-cluster antikink $|2n\rangle _1$, 
(f) and (g)  5-cluster antikinks $|2n\rangle _2$ and $|2n\rangle _3$, 
(h) 9-cluster antikink $|2n\rangle _4$, (i) singlet dimer $|O\rangle$.}}
\end{center}
\end{figure}

The AFM $J_1$-$J_2$ sawtooth spin-$\frac{1}{2}$ Heisenberg chain is depicted in Fig.\ref{sawtooth} (a). 
The model is defined by the Hamiltonian,  
\bea 
H\!=\!\sum_{i=1}^{N}H_i,\; H_i\!=\!J_1\left(h_{2i-1,2i}+h_{2i,2i+1}\right)+J_2 \;h_{2i-1,2i+1},\;
h_{m,n}=S_{m}^xS_{n}^x+S_{m}^yS_{n}^y+\Delta S_{m}^zS_{n}^z.
\label{ham}
\eea
$N$ is the total number of triangles in the chain. $J_1$ and $J_2$ are the 
nearest neighbour (NN) and the next nearest neighbour (NNN) AFM exchange strengths, 
respectively. $\vec S_m$ is the spin-1/2 operator at site $m$. 
Now and henceforth $\Delta$ is the anisotropic parameter, $0\leq \Delta\leq 1$. 
This model becomes a single AFM Heisenberg chain when $J_2/J_1=0$, where the low-energy 
excitations are spinons for $0\leq \Delta\leq 1$ \cite{Takhtajan},  
or domain-walls, for $\Delta>1$ \cite{McCoy,Villain,Ghosh}. 
In the other extreme limit, $J_2/J_1\rightarrow \infty$,  
the model decouples into a single AFM Heisenberg chain along the base 
of the triangles due to the strong exchange interactions among the NNN spins 
and a number of nearly free spins 
sit on the vertices of every triangle.
The model at the symmetric point, $J_1=J_2$ and at the isotropic point 
$\Delta=1$ has been studied extensively \cite{Sen-Shastry,Nakamura,Kubo}. 
The asymmetric sawtooth model where  $J_1\neq J_2$ is 
numerically investigated at the isotropic point \cite{Blundell}. 

The model has the global $U(1)$ symmetry 
since the $z$-component of the total spin,  
$S^z_{\rm T}$, is always a good quantum number. 
The ground state is doubly degenerate when total number of sites 
is even and the periodic boundary condition is considered. 
These two ground states $|G_1\rangle$ and $|G_2\rangle$, shown in Fig.\ref{sawtooth} (b) 
and (c), respectively, are also the ground states of MG model \cite{Majumdar-Ghosh}. 
They are the product of singlet dimers, $|O\rangle$ formed over 
the alternate NN bonds constituted by the spin-1/2 states at their two ends.  
So, the two ground states would be expressed as $|G_j\rangle=\prod _i^N |O_j\rangle _i$, where  
$|O_j\rangle _i\!=\!(|+\rangle _{_{2i+j-2}}\,|-\rangle _{_{2i+j-1}}\!\!-\!
|-\rangle _{_{2i+j-2}}\,|+\rangle _{_{2i+j-1}})/\sqrt 2$, 
$S_i^z|\pm\rangle_i\!=\!\pm \frac{1}{2}|\pm\rangle_i$,  
and  $j=1,2$. $|G_1\rangle$ and $|G_2\rangle$ are connected by the 
lattice translation of one unit along the NN bonds and orthogonal 
when $N\rightarrow\infty$.
On the other hand, for the open chain with odd number of sites 
ground state is $2(N+1)$-fold degenerate. Among them $2(N-1)$ number of states 
are known as kink in which the free spin-1/2 state separates 
the $|G_1\rangle$ on the left and $|G_2\rangle$ on the right \cite{Nakamura,Sen-Shastry}. 
One such state, $|K\rangle$, is shown in Fig.\ref{sawtooth} (d). 
In the remaining 
four states free spin-1/2 state remains at one of the end.  

It can be shown that all those states described above are 
the exact eigenstates of the Hamiltonian $H$ when $J_1=J_2$ 
for any values of $\Delta$ having the total 
ground state energy, $E_{\rm G}=-\frac{N}{2}(1+\frac{\Delta}{2})J_1$. 
This result can be proved in the following way. 
For the $i$-th triangle, $H_i$ can be expressed as $H_i=J_1(H_i^*+(\Delta-1)H_i^z)$, when  $J_1=J_2$, where 
$H_i^*=\vec S_{2i-1}\cdot \vec S_{2i}+\vec S_{2i}\cdot \vec S_{2i+1}+\vec S_{2i+1}\cdot \vec S_{2i+1}$ and 
$H_i^z= S_{2i-1}^z S_{2i}^z+S_{2i}^z S_{2i+1}^z+ S_{2i+1}^zS_{2i+1}^z$. $H_i^*$ and $H_i^z$ commute with each other. 
It has been shown that for the $i$-th triangle constituted by the three spins, $S_{2i-1}$, $S_{2i}$ and $S_{2i+1}$, 
$H_i^*=\frac{3}{2}[\mathcal P_{i} -\frac{1}{2}]$, where $\mathcal P_{i}$ is the projection operator 
such that $\mathcal P_{i} |S_i\!=\!\frac{1}{2}\rangle\!=\!0|S_i\!=\!\frac{1}{2}\rangle$ and 
$\mathcal P_{i} |S_i\!=\!\frac{3}{2}\rangle\!=\!1|S_i\!=\!\frac{3}{2}\rangle$, in which 
$\vec S_i\!=\!(\vec S_{2i-1}\!+\! \vec S_{2i} \!+ \!\vec S_{2i+1})$, is 
total spin vector for the $i$-th triangle \cite{Sen-Shastry}. 
Therefore, $H_i^*|S_i\!=\!\frac{1}{2}\rangle\!=\!-\frac{3}{4}|S_i\!=\!\frac{1}{2}\rangle$. 
$|S_i\!=\!\frac{1}{2}\rangle$ state arises in case only when a singlet is realized in the $i$-th 
triangle which is also the eigenstate of $H_i^z$  having the lowest eigenvalue, $-1/4$. 
Hence $|S_i\!=\!\frac{1}{2}\rangle$ is the lowest energy eigenstate of 
both the commuting Hamiltonians 
 $H_i^*$ and $H_i^z$ having eigenvalues $-3/4$ and  $-1/4$, 
respectively and thus gives rise to the lowest eigenvalue, $-\frac{1}{2}(1+\frac{\Delta}{2})J_1$ of the 
Hamiltonian $H_i$, which on the other hand becomes equal to the ground state energy per triangle, $E_G/N$.  
When $J_1\neq J_2$, those state are not the eigenstates of $H$, but 
the expectation value $\langle H\rangle=E_{\rm G}$. Since all kinks have the same energy 
and equal to the ground state energy, they are dispersionless for this sawtooth model. 
However, in case of MG model, kinks form dispersive 
mode and yield spin-gap \cite{Caspers}. 

For the open chain low energy excitations for this model are formed by antikinks and those are not 
the exact eigenstates even when $J_1=J_2$ \cite{Nakamura,Sen-Shastry}. 
In contrast to the kink, antikink separates 
the $|G_2\rangle$ on the left and $|G_1\rangle$ on the right. In the antikinks, 
$|G_1\rangle$ and $|G_2\rangle$ states are separated by the odd number of sites. 
Therefore, $m$-cluster antikinks could be formed where $m=1,3,5,\cdots$. 
Those antikinks are not orthogonal to each other.
It has been shown that 3-cluster and  7-cluster antikinks decompose into 1-cluster 
and 5-cluster antikinks, respectively, and this decomposition procedure 
is still applicable to much higher cluster antikinks \cite{Caspers,Sen-Shastry}. 
So, in this article, dispersion relations of 1-, 5- and 9-cluster antikinks 
will be obtained since more than 9-cluster antikinks are likely to contribute less  
in the low energy excitations. Further, unlike the 1- and 9-cluster antikinks, 
there is two different dimer orientations for 
5-cluster antikinks and those are connected by the mirror symmetry about the 
central site, $2n$.  1-, 5- and 9-cluster antikinks are noted as $|2n\rangle _1$, 
\{$|2n\rangle _2$,\;$|2n\rangle _3$\} and $|2n\rangle _4$ and they are 
shown in Fig.\ref{sawtooth} (e), \{(f), (g)\} and (h), respectively.  
The value of net spin of all the kink and antikink states is 1/2 for the open chain. 

\section{Antikink dispersions and spin-gap}
\label{A-D}
To obtain the antikink dispersions, linear superposition of 
1-, 5- and 9-cluster antikinks are obtained separately for a definite 
momentum $k$ when $N\rightarrow \infty$. In this case, open chain with 
odd number of sites is considered. Those momentum eigenstates 
of antikinks are written as, 
\be
|k\rangle_i=\frac{1}{\sqrt N}\sum_{n=0}^N\,e^{ikn}\;|2n\rangle_i,\quad {\rm where}\quad i=1,2,3,4.
\ee
The states, $|k\rangle_i$ are not normalized and rather the overlaps 
$_i\langle k|k\rangle_i $ are found to be a functions of $k$. 
For example, in case of 1-cluster antikinks, the matrix elements, $_1\langle 2n|2m\rangle_1=(-1/2)^{|n-m|}$.   
So, the value of $_1\langle k|k\rangle_1 =f(\cos{k})$ when $N\rightarrow \infty$, 
where $f(x)=3/(5+4x)$ \cite{Sen-Shastry}. 
Similar types of matrix elements for 5- and 9-cluster antikinks are 
\bea
{}_2\langle 2n|2m\rangle_2 &=&
3/2\,\delta_{n,m}-3/8\left(\delta_{n,m+1}+\delta_{n,m-1} \right)
+(-1/2)^{|n-m|+1},\nonumber\\ 
{}_4\langle 2n|2m\rangle_4 &=&3/4\,\delta_{n,m}
+3/32\left(\delta_{n,m+1}+\delta_{n,m-1} +\delta_{n,m+3}+\delta_{n,m-3} \right)\nonumber\\
&&-3/16\left(\delta_{n,m+2}+\delta_{n,m-2} \right)
+(-1/2)^{|n-m|+2}, \nonumber
\eea  
and ${}_2\langle 2n|2m\rangle_2 ={}_3\langle 2n|2m\rangle_3$.
Those matrix elements 
lead to the following expressions of overlaps when $N\rightarrow \infty$, 
\bea
{}_2\langle k|k\rangle_2 &=&
\frac{3}{2}-\frac{3}{4}\cos{k}-\frac{1}{2}f(\cos{k}), \nonumber\\
{}_4\langle k|k\rangle_4 &=&\frac{3}{4}+\frac{3}{16}\left(\cos{k}+\cos{3k}\right)
-\frac{3}{8}\cos{2k}+\frac{1}{4}f(\cos{k}), \nonumber
\eea
and ${}_2\langle k|k\rangle_2 ={}_3\langle k|k\rangle_3$.
In order to derive the dispersion relations, required matrix elements of $H$ are 
obtained. Expressions of those matrix elements are available in Appendix \ref{A1}.
1-, 5- and 9-cluster antikink dispersion relations, 
$\omega_i(k)={}_i\langle k|H|k\rangle_i/{}_i\langle k|k\rangle_i -E_{\rm G},\;i=1,2,4$, are therefore 
given by,
\bea
\omega_1(k)&=&\frac{g(\Delta)}{2\;{}_1\langle k|k\rangle_1}(J_1-(J_1\!-\!J_2)h(\cos{k})),\nonumber \\
\omega_2(k)&=&\frac{g(\Delta)}{{}_2\langle k|k\rangle_2}\bigg[J_1\left(1-\frac{1}{4}\cos{2k}\right)
+(J_1\!-\!J_2)\bigg(2-\frac{3}{2}\cos{k}
+\frac{1}{4}\cos{2k}-\frac{3}{2}f(\cos{k})+\frac{1}{4}h(\cos{k})\bigg)\bigg], \nonumber \\
\omega_4(k)&=&\frac{g(\Delta)}{{}_4\langle k|k\rangle_4}\bigg[J_1\bigg(\frac{3}{2}-\frac{1}{4}\cos{2k}
+\frac{1}{16}\cos{4k}\bigg)+(J_1\!-\!J_2)\bigg(-\frac{7}{8} 
+\frac{9}{8}\cos{k}-\frac{15}{16}\cos{2k}+\frac{1}{2}\cos{3k}\nonumber \\
&&-\frac{1}{8}\cos{4k}
-\frac{11}{8}f(\cos{k})-\frac{1}{8}h(\cos{k})\bigg)\bigg],  \nonumber 
\eea
where $g(\Delta)=1+\Delta/2$ and $h(x)=4(4+5x)/(25+40x+16x^2)$. 
 \begin{figure}[h]
\begin{center}
\psfrag{j1}{$J_1$}
\psfrag{j2}{\text{$\frac{J_2}{J_1}$}}
\psfrag{w1}{\text{$\omega_1(k)/J_1$}}
\psfrag{w2}{\text{$\omega_2(k)/J_1$}}
\psfrag{w4}{\text{$\omega_4(k)/J_1$}}
\psfrag{a}{\text{\scriptsize{(a)}}}\psfrag{b}{\text{\scriptsize{(b)}}}
\psfrag{c}{\text{\scriptsize{(c)}}}\psfrag{d}{\text{\scriptsize{(d)}}}
\psfrag{e}{\text{\scriptsize{(e)}}}\psfrag{f}{\text{\scriptsize{(f)}}}
\psfrag{g}{\text{\scriptsize{(g)}}}\psfrag{h}{\text{\scriptsize{(h)}}}
\psfrag{2n}{\text{\scriptsize{$2n$}}}
\psfrag{a1}{\text{\scriptsize{$|2n\rangle _1$}}}
\psfrag{a2}{\text{\scriptsize{$|2n\rangle _2$}}}
\psfrag{a3}{\text{\scriptsize{$|2n\rangle _3$}}}
\psfrag{a4}{\text{\scriptsize{$|2n\rangle _4$}}}
\psfrag{g1}{\text{\scriptsize{$|G_1\rangle$}}}
\psfrag{g2}{\text{\scriptsize{$|G_2\rangle$}}}
\psfrag{k}{\text{$k$}}
\psfrag{0.0}{\text{\scriptsize{$0.0$}}}
\psfrag{0.5}{\text{\scriptsize{$0.5$}}}
\psfrag{1.0}{\text{\scriptsize{$1.0$}}}
\psfrag{1.5}{\text{\scriptsize{$1.5$}}}
\psfrag{2.0}{\text{\scriptsize{$2.0$}}}
\psfrag{2.5}{\text{\scriptsize{$2.5$}}}
\psfrag{3.0}{\text{\scriptsize{$3.0$}}}
\psfrag{0.00}{\text{\scriptsize{$0.0$}}}
\psfrag{1.57}{\text{\scriptsize{$\pi/2$}}}
\psfrag{4.71}{\text{\scriptsize{$3\pi/2$}}}
\psfrag{3.14}{\text{\scriptsize{$\pi$}}}
\psfrag{6.28}{\text{\scriptsize{$2\pi$}}}
\psfrag{0.60}{\text{\scriptsize{$0.6$}}}
\psfrag{0.80}{\text{\scriptsize{$0.8$}}}
\psfrag{1.00}{\text{\scriptsize{$1.0$}}}
\psfrag{1.20}{\text{\scriptsize{$1.2$}}}
\psfrag{1.40}{\text{\scriptsize{$1.4$}}}
\psfrag{1.60}{\text{\scriptsize{$1.6$}}}
\psfrag{a2}{\text{\scriptsize{$|2n\rangle _2$}}}
\psfrag{s}{\text{\scriptsize{$\equiv\frac{1}{\sqrt 2}(\us\ds-\ds\us)$}}}
\includegraphics[scale=0.80]{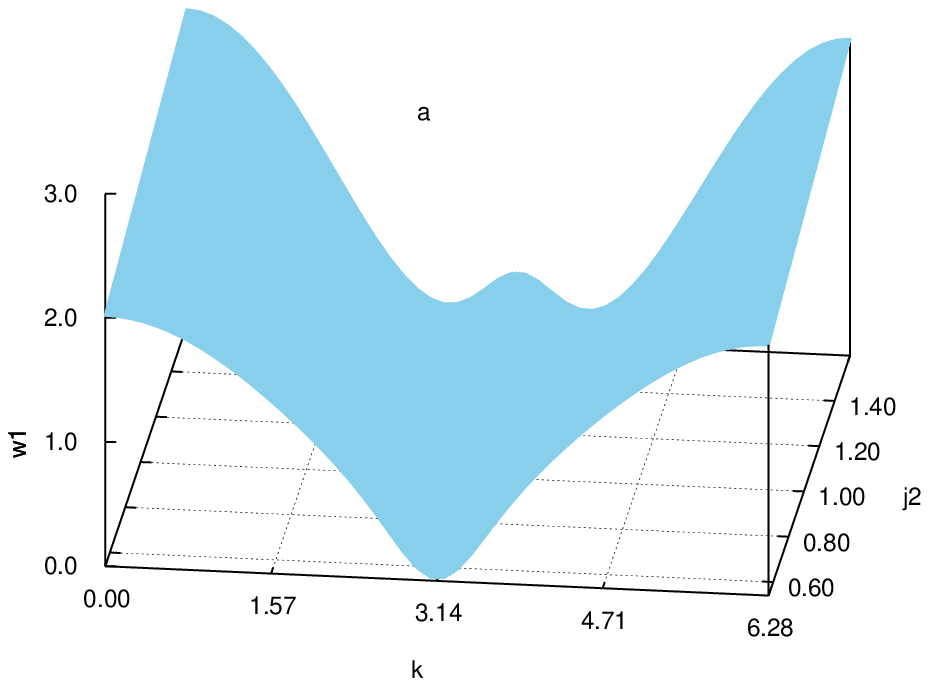}
\vskip -.6 in
\includegraphics[scale=0.80]{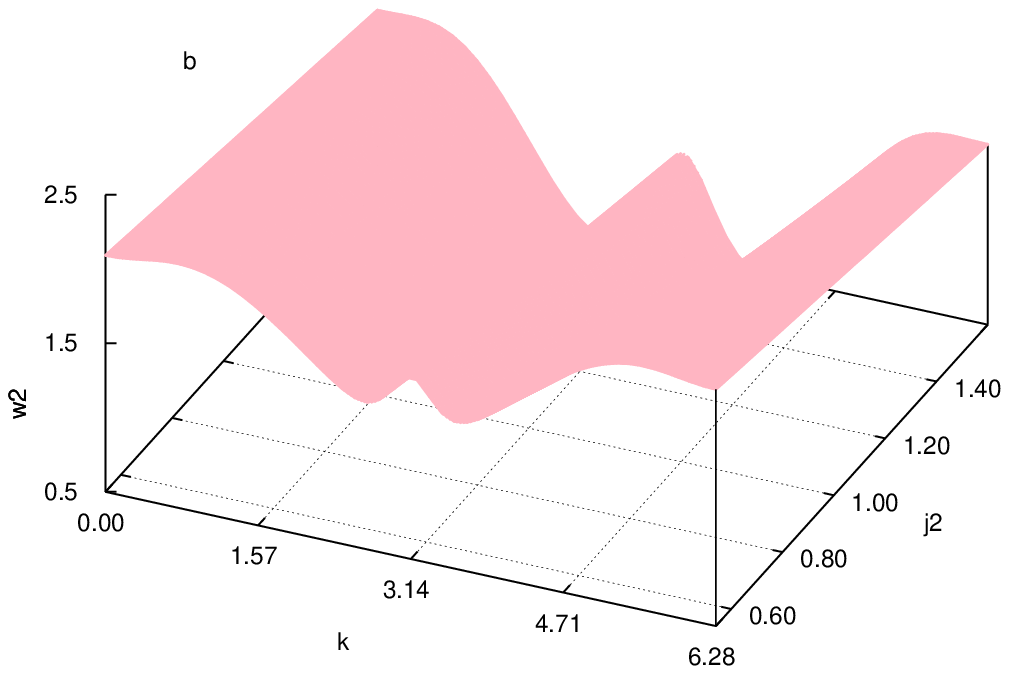}
\vskip -.5 in
\includegraphics[scale=0.80]{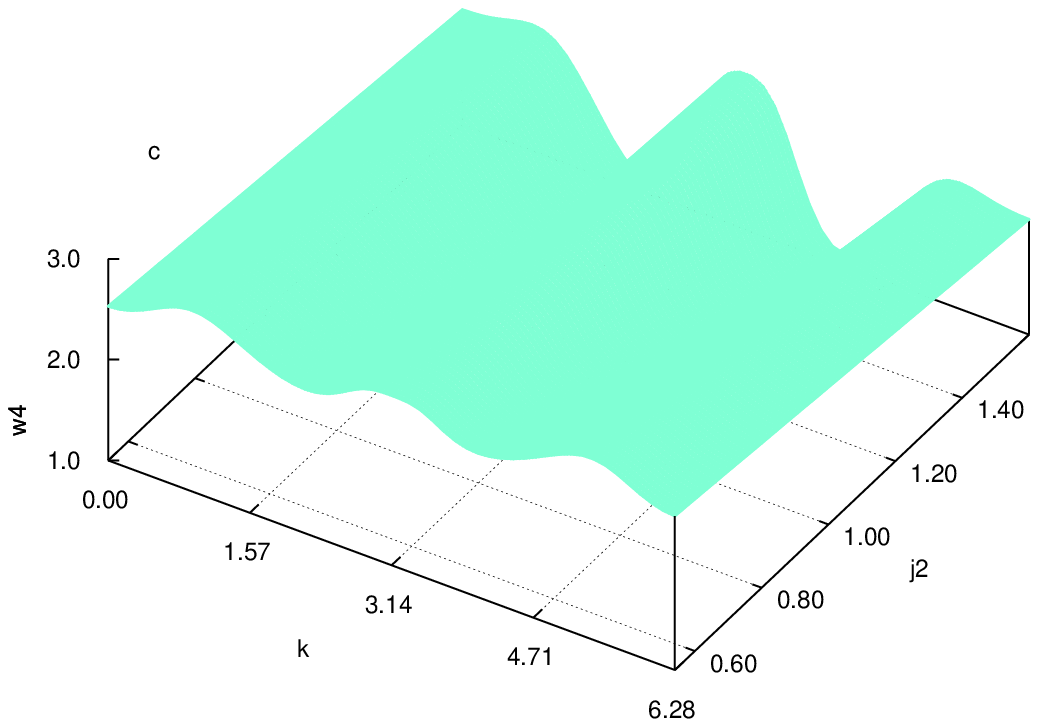}
\caption{\label{antikink-dispersion}{(Color Online) 
Antikink dispersions for $\Delta=1.0$: (a) $\omega_1(k)$, 
(b) $\omega_2(k)$ and (c) $\omega_4(k)$.}}
\end{center}
\end{figure}
Variation of $\omega_i(k)$ with respect to $J_2/J_1$ in the first Brilloiun zone 
are shown in Fig.\ref{antikink-dispersion}. The range of $J_2/J_1$ is bounded within 
 $0.542<J_2/J_1<1.598$, because of the fact that spin-gap 
vanishes beyond those limits and thus the 
antikinks are no longer a valid excitations. 
$\omega_1(k)$ has a maximum at $k=0$ and a 
minimum at $k=\pi$ as long as $J_2\leq J_1$ 
but an additional peak appears at $k=\pi$ when $J_2> J_1$. So, the number of minima in 
$\omega_1(k)$ is two and symmetric around $k=\pi$ when $J_2> J_1$.
Number of maxima for $\omega_2(k)$ 
and $\omega_4(k)$ are three in which one maxima and one minima are always 
at $k=\pi$ and $k=0$, respectively. The minimum value of $\omega_i(k)$ increases 
from 1- to 9-cluster antikinks. Therefore 
higher cluster antikinks contribute less to the low-temperature dynamics. 

\begin{figure}[h]
\begin{center}
\psfrag{delta}{$\Delta$}
\psfrag{s-g}{$E_{\rm Gap}/J_1$}
\psfrag{j2}{\text{$\frac{J_2}{J_1}$}}
\psfrag{wab}{\text{$\omega_m(k)$}}
\psfrag{a}{\text{\scriptsize{(a)}}}\psfrag{b}{\text{\scriptsize{(b)}}}
\psfrag{k}{\text{$k$}}
\psfrag{0.0}{\text{\scriptsize{$0.0$}}}
\psfrag{0.5}{\text{\scriptsize{$0.5$}}}
\psfrag{1.0}{\text{\scriptsize{$1.0$}}}
\psfrag{1.5}{\text{\scriptsize{$1.5$}}}
\psfrag{2.0}{\text{\scriptsize{$2.0$}}}
\psfrag{2.5}{\text{\scriptsize{$2.5$}}}
\psfrag{3.0}{\text{\scriptsize{$3.0$}}}
\psfrag{0.00}{\text{\scriptsize{$0.0$}}}
\psfrag{1.57}{\text{\scriptsize{$\pi/2$}}}
\psfrag{4.71}{\text{\scriptsize{$3\pi/2$}}}
\psfrag{3.14}{\text{\scriptsize{$\pi$}}}
\psfrag{6.28}{\text{\scriptsize{$2\pi$}}}
\psfrag{0.60}{\text{\scriptsize{$0.6$}}}
\psfrag{0.80}{\text{\scriptsize{$0.8$}}}
\psfrag{1.00}{\text{\scriptsize{$1.0$}}}
\psfrag{1.20}{\text{\scriptsize{$1.2$}}}
\psfrag{1.40}{\text{\scriptsize{$1.4$}}}
\psfrag{1.60}{\text{\scriptsize{$1.6$}}}
\psfrag{0.6}{\text{\scriptsize{$0.6$}}}
\psfrag{0.1}{\text{\scriptsize{$0.1$}}}
\psfrag{0.2}{\text{\scriptsize{$0.2$}}}
\psfrag{0.3}{\text{\scriptsize{$0.3$}}}
\psfrag{0.4}{\text{\scriptsize{$0.4$}}}
\psfrag{0.8}{\text{\scriptsize{$0.8$}}}
\psfrag{1.2}{\text{\scriptsize{$1.2$}}}
\psfrag{1.4}{\text{\scriptsize{$1.4$}}}
\psfrag{1.6}{\text{\scriptsize{$1.6$}}}
\includegraphics[scale=0.80]{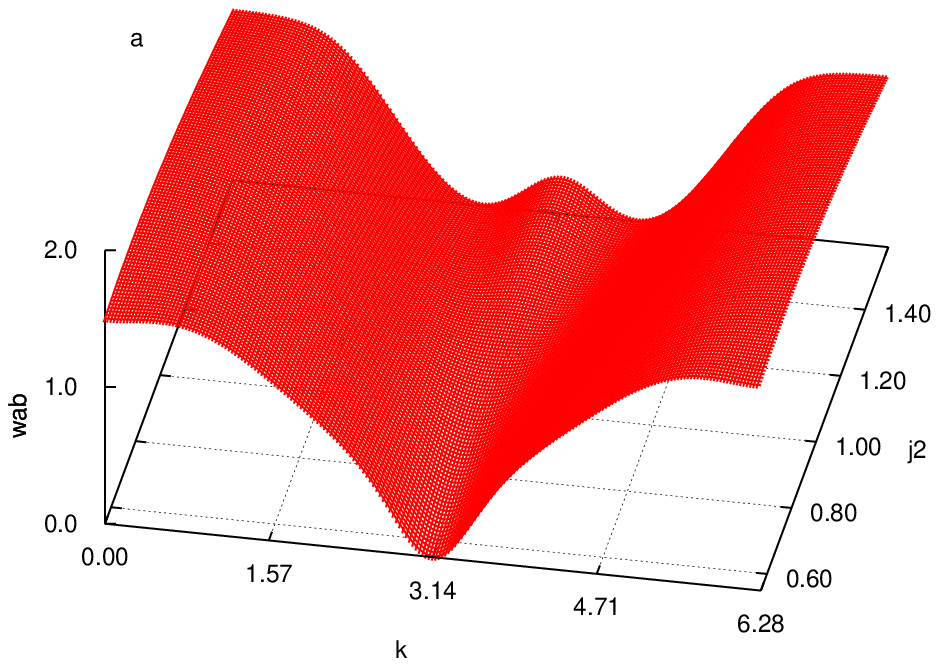}
\vskip -.7 in
\includegraphics[scale=0.80]{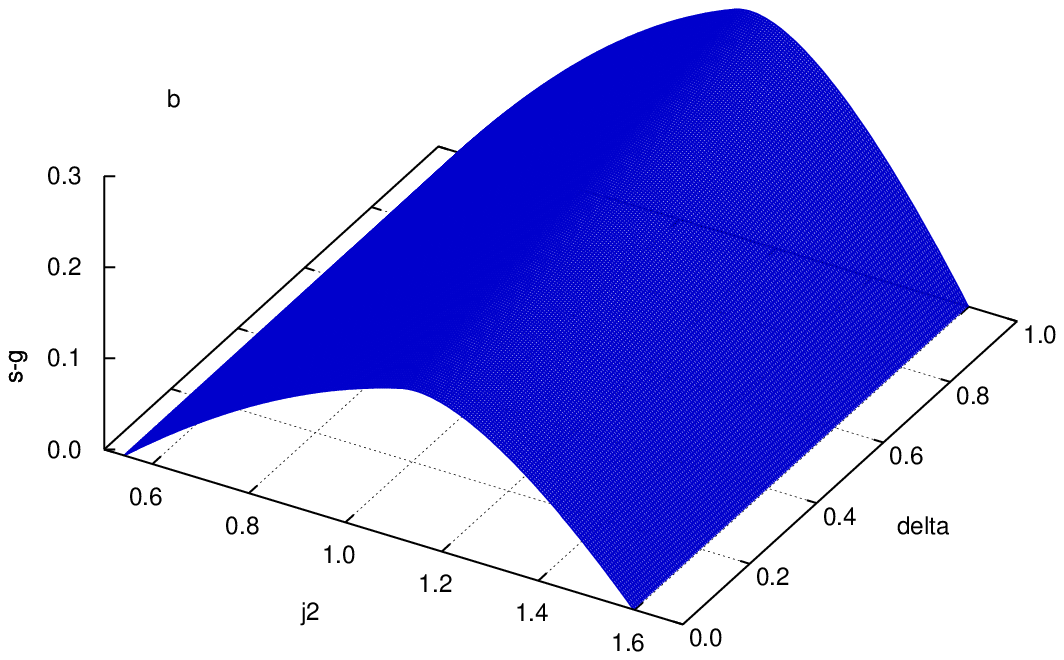}
\caption{\label{minimized-dispersion}{(Color Online) 
(a) Antikink dispersion,  $\omega_m(k)$ when $\Delta=1.0$, 
and (b) variation of $E_{\rm Gap}$ with $J_2/J_1$ and $\Delta$. }}
\end{center}
\end{figure}

Since those antikinks are not orthogonal to each other,
 a variational approach has been adopted to estimate the spin-gap 
in the thermodynamic limit \cite{Sen-Shastry}. The variational state is defined as 
\be
|k\rangle=\frac{1}{\sqrt N}\sum_{n=0}^N\,e^{ikn}\;\left[|2n\rangle_1+a(|2n\rangle_2+|2n\rangle_3)+b|2n\rangle_4\right].
\ee
which is again a momentum eigenstate. The variational parameters $a$ and $b$ are considered as real. 
Since the dispersions of two 5-cluster antikinks are the same, the variational parameter 
for both of them is $a$. The contribution of antikinks spread over more 
than nine sites would be negligible in the 
estimation of spin-gap, and thus not considered. The variational dispersion relation,  
$\omega_k(a,b)=\langle k|H|k\rangle/\langle k|k\rangle-E_{\rm G}$, will be minimized with respect to 
both $a$ and $b$. The additional matrix elements for obtaining $\langle k|H|k\rangle$ and 
$\langle k|k\rangle$ are available in Appendix \ref{A2}. 
$\omega(k)$ has been minimized numerically by using simplex minimizing 
procedure \cite{Nelder_Mead}. 
The minimized dispersion, $\omega_m(k)$, has been plotted 
with respect to $J_2/J_1$ in the first Brilloiun zone (Fig \ref{minimized-dispersion}(a)). 
It looks similar to the 1-cluster dispersion, $\omega_1(k)$ since the contribution of higher cluster 
antikinks is very less.  
$E_{\rm Gap}$ has been estimated by minimizing $\omega_m(k)$ with respect to $k$.  
Variation of $E_{\rm Gap}$  with respect to 
$J_2/J_1$ and $\Delta$ is shown in Fig \ref{minimized-dispersion} (b). 
$E_{\rm Gap}$ has a peak at the point $J_2/J_1=1$ for any values of $\Delta$, 
and the curve is not symmetric around this peak. 
The value of this peak decreases with the decrease of $\Delta$. 
$E_{\rm Gap}$ is non-zero as long as $0.542\leq J_2/J_1\leq 1.598$, 
irrespective of the value of 
$\Delta$.  $E_{\rm Gap}= 0.2192 J_1$ when $J_2/J_1=1$ and $\Delta=1$ where  
the minimum of $\omega(k)$ appears at $a=-0.2808,\;b=0$ and $k=\pi$. 
At this particular point, $E_{\rm Gap}$ becomes equal to that estimated before 
by considering 1- and 5-cluster antikinks \cite{Sen-Shastry},  
because of the fact that here 9-cluster antikink contributes nothing. 
However, contribution of all the three clusters 
in the $\omega_m(k)$ is found in the rest of the parameter space. 
When $E_{\rm Gap}$ vanishes at $J_2/J_1=0.542$, $a=0.1680$ and $b=0.1738$, 
whereas at $J_2/J_1=1.598$, $a=-0.5976$ and $b=0.2821$. 
$E_{\rm Gap}= 0.1461 J_1$ when $J_2/J_1=1$ and $\Delta=0$.
Minimum of $\omega_m(k)$ is found when $k=\pi$ for $J_2\leq J_1$ 
and $k\approx\pi$ for $J_2>J_1$, irrespective of the value of $\Delta$. 
Beyond the region $0.542\leq J_2/J_1\leq 1.598$, $\omega_m(k)$ becomes 
negative which means that antikinks are no longer valid excitations. 
Here spinons may constitute the low energy excitations instead of antikinks 
particularly when $J_2/J_1<0.542$. 
\section{Exact diagonalization results}
\label{E-D} 
In order to obtain the value of $E_{\rm Gap}$ numerically, a chain containing 
$N$ triangles is considered.  
The Lanczos exact diagonalization algorithm is the most useful in this case 
since only two lowest eigenvalues are required for the 
estimation of $E_{\rm Gap}$. 
As the ground state always lies in the $S_{\rm T}^z=0$ sector irrespective 
of the values of any parameters and degeneracy of it, 
the Hamiltonian has been diagonalized in the  $S_{\rm T}^z=0$ sub-space 
for the estimation of ground state energy, $E_{\rm G}$. 
Owing to the translational symmetry of the Hamiltonian by one triangle 
or two lattice sites, momentum wave vectors, $k$ has now discrete values. 
In the $S_{\rm T}^z=0$ subspace, an additional symmetry 
composing of spin inversion in every site 
is considered for further reduction of the Hilbert space. 
Let $q$ be the momentum wave vector of this spin inversion symmetry, 
which may obtain value either 0 or $\pi$.  
The first excited state always appear in the $S_{\rm T}^z=1$ sector 
when it is nondegenerate. In case of multiple degeneracy, the additional states 
may appear in the $S_{\rm T}^z=0$ sector depending on the values of the 
parameters, $J_2$, $\Delta$ and $N$. To estimate the 
first excited state energy, $E_{\rm F}$, the Hamiltonian 
is thus diagonalized in the $S_{\rm T}^z=1$ sub-space. 
Since this sub-space lacks the spin inversion symmetry, the 
eigenstates are defined only by the values of $k$. 
Finally, including those symmetries in   
this computational procedure,  
sawtooth chain up to $N=14$, or 28 sites have been considered. 
The extrapolated spin-gap is defined as 
$E_{\rm Gap}=\lim_{N\rightarrow \infty}
[E_{\rm F}(N,S_{\rm T}^z=1)-E_{\rm G}(N,S_{\rm T}^z=0)]$, and that  
is obtained by using the Vanden-Broeck-Schwartz algorithm \cite{VBS}. 

When $J_2\!=\!J_1$ and $0\!\leq\!\Delta\!\leq\!1$, 
ground state is doubly-degenerate. 
Both the states have the same momenta, $k\!=\!0,\,q\!=\!0$ when $N$ is even. 
But when $N$ is odd, those states are defined by different set of momenta, 
$k\!=\!0,\,q\!=\!0$ and $k\!=\!0,\,q\!=\!\pi$. 
The first excited state is always doubly degenerate 
when $\Delta=1$. They both lie in the  $S_{\rm T}^z=1$ sector when $N$ is odd, and otherwise  
 in two different, $S_{\rm T}^z=0,1$ sectors when $N$ is even. 
The non-dispersive character of the degenerate first excited state when 
$\Delta=1$ is reported before \cite{Kubo,Blundell}. 
For $0\leq\Delta<1$, first excited state 
is non-degenerate for even $N$ but doubly degenerate for odd $N$. Whatever may be the case, 
one of the first excited state is always appear with the value $k=\pi$, for even $N$. 

When $J_2\!<\!J_1$ and $0\!\leq\!\Delta\!\leq\!1$, both ground 
and first excited states 
are non-degenerate. The unique ground state always 
belongs to the $S_{\rm T}^z=0$ sector with $k\!=\!0,\,q\!=\!0$ when $N$ is even. 
On the other hand, the first excited state is found to undergo  
a crossover from $S_{\rm T}^z=0$ to $S_{\rm T}^z=1$ sector with the 
decrease of $J_2$. 
The momentum of this state has opted the value $k=0$, when it 
appears in the $S_{\rm T}^z=1$ sector. 
The value of $J_2$ for the crossover point depends on the value of 
$\Delta$ and $N$. 
For $J_2\!>\!J_1$ and $0\!\leq\!\Delta\!\leq\!1$, again both ground 
and first excited states 
are non-degenerate. In this case, first excited state has the value 
$k=\pi$, and it appears in the $S_{\rm T}^z=1$ sector. 

\begin{figure}[h]
\begin{center}
\psfrag{delta}{$\Delta$}
\psfrag{s-g}{$E_{\rm Gap}/J_1$}
\psfrag{j2}{\text{$\frac{J_2}{J_1}$}}
\psfrag{c}{\text{\scriptsize{}}}
\psfrag{k}{\text{$k$}}
\psfrag{0.0}{\text{\scriptsize{$0.0$}}}
\psfrag{0.5}{\text{\scriptsize{$0.5$}}}
\psfrag{1.0}{\text{\scriptsize{$1.0$}}}
\psfrag{1.5}{\text{\scriptsize{$1.5$}}}
\psfrag{2.0}{\text{\scriptsize{$2.0$}}}
\psfrag{2.5}{\text{\scriptsize{$2.5$}}}
\psfrag{3.0}{\text{\scriptsize{$3.0$}}}
\psfrag{0.00}{\text{\scriptsize{$0.0$}}}
\psfrag{1.57}{\text{\scriptsize{$\pi/2$}}}
\psfrag{4.71}{\text{\scriptsize{$3\pi/2$}}}
\psfrag{3.14}{\text{\scriptsize{$\pi$}}}
\psfrag{6.28}{\text{\scriptsize{$2\pi$}}}
\psfrag{0.60}{\text{\scriptsize{$0.6$}}}
\psfrag{0.80}{\text{\scriptsize{$0.8$}}}
\psfrag{1.00}{\text{\scriptsize{$1.0$}}}
\psfrag{1.20}{\text{\scriptsize{$1.2$}}}
\psfrag{1.40}{\text{\scriptsize{$1.4$}}}
\psfrag{1.60}{\text{\scriptsize{$1.6$}}}
\psfrag{0.6}{\text{\scriptsize{$0.6$}}}
\psfrag{0.1}{\text{\scriptsize{$0.1$}}}
\psfrag{0.2}{\text{\scriptsize{$0.2$}}}
\psfrag{0.3}{\text{\scriptsize{$0.3$}}}
\psfrag{0.4}{\text{\scriptsize{$0.4$}}}
\psfrag{0.8}{\text{\scriptsize{$0.8$}}}
\psfrag{1.2}{\text{\scriptsize{$1.2$}}}
\psfrag{1.4}{\text{\scriptsize{$1.4$}}}
\psfrag{1.6}{\text{\scriptsize{$1.6$}}}
\includegraphics[scale=0.80]{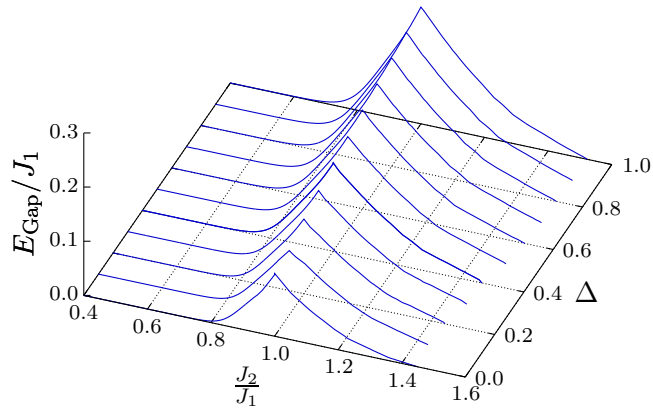}
\caption{\label{extrapolated_spin-gap}{(Color Online) 
Variation of numerically extrapolated spin-gap, $E_{\rm Gap}/J_1$, with $\Delta$ 
and $J_2/J_1$.}}
\end{center}
\end{figure}

Variation of estimated $E_{\rm Gap}$ has been shown in 
Fig.\ref{extrapolated_spin-gap}. Its value is lower than that 
obtained from antikink dispersion in most of the regions. 
Although there is a overall similarity  
between the two estimations regarding its variation with 
respect to $J_2/J_1$ and $\Delta$, the peak is more sharp in case of 
numerical study. Other differences include the range of $J_2/J_1$ where 
the spin-gap is non-zero. For $\Delta=1$, $E_{\rm Gap}\neq 0$ when 
$0.487\leq J_2/J_1 \leq 1.531$, which is the same to the previous estimation 
\cite{Blundell}. The non-zero spin-gap region shrinks with the decrease of 
$\Delta$. On the other hand, antikink dispersion determines non-zero spin-gap 
in the region, $0.542\leq J_2/J_1\leq 1.598$, irrespective of the values of 
$\Delta$. This disagreement may attribute to the differences of those 
two approaches. In the variational method, solitary antikink excitations 
in the open chain are considered. While in the exact diagonalization, 
periodic chain is considered where kink-antikink pair states constitute the 
low-energy excitations \cite{Nakamura}. 
$E_{\rm Gap}=0.215J_1$, when $J_2/J_1=1$ and  $\Delta=1$, 
which is again the same to the previous estimations \cite{Nakamura,Blundell}. 
The value of $E_{\rm Gap}$ predicted from antikink dispersion
at this point is $0.2192J_1$, which is marginally higher. 
The value of $E_{\rm Gap}$ decreases with the decrease of $\Delta$, 
which is similar to the results obtained in the variational method. 
For $J_2/J_1=1$ and  $\Delta=0$, the numerical value of $E_{\rm Gap}=0.116J_1$, 
while the variational estimation is $E_{\rm Gap}=0.1461J_1$. The difference between 
numerical and variational estimations increases with the decrease of 
$\Delta$ for fixed value of  $J_2/J_1$.

\begin{figure}[h]
\begin{center}
\psfrag{T}{\text{$k_{\rm B}T/J_1$}}
\psfrag{s-p}{\text{$C_{\rm V}/J_1$}}
\psfrag{sus}{\text{$\chi/J_1$}}
\psfrag{j2}{\text{$\frac{J_2}{J_1}$}}
\psfrag{del1.0}{\hskip .3 cm\text{$\Delta=1.0$}}
\psfrag{a}{\hskip -0.05 cm \text{{(a)}}}
\psfrag{b}{\text{{(b)}}}
\psfrag{k}{\text{$k$}}
\psfrag{0.0}{\text{\scriptsize{$0.0$}}}
\psfrag{0.5}{\text{\scriptsize{$0.5$}}}
\psfrag{1.0}{\text{\scriptsize{$1.0$}}}
\psfrag{1.5}{\text{\scriptsize{$1.5$}}}
\psfrag{2.0}{\text{\scriptsize{$2.0$}}}
\psfrag{2.5}{\text{\scriptsize{$2.5$}}}
\psfrag{0.00}{\text{\scriptsize{$0.00$}}}
\psfrag{0.05}{\text{\scriptsize{$0.05$}}}
\psfrag{0.10}{\text{\scriptsize{$0.10$}}}
\psfrag{0.15}{\text{\scriptsize{$0.15$}}}
\psfrag{0.20}{\text{\scriptsize{$0.20$}}}
\psfrag{0.25}{\text{\scriptsize{$0.25$}}}
\psfrag{0.30}{\text{\scriptsize{$0.30$}}}
\psfrag{0.35}{\text{\scriptsize{$0.35$}}}
\includegraphics[scale=1.00]{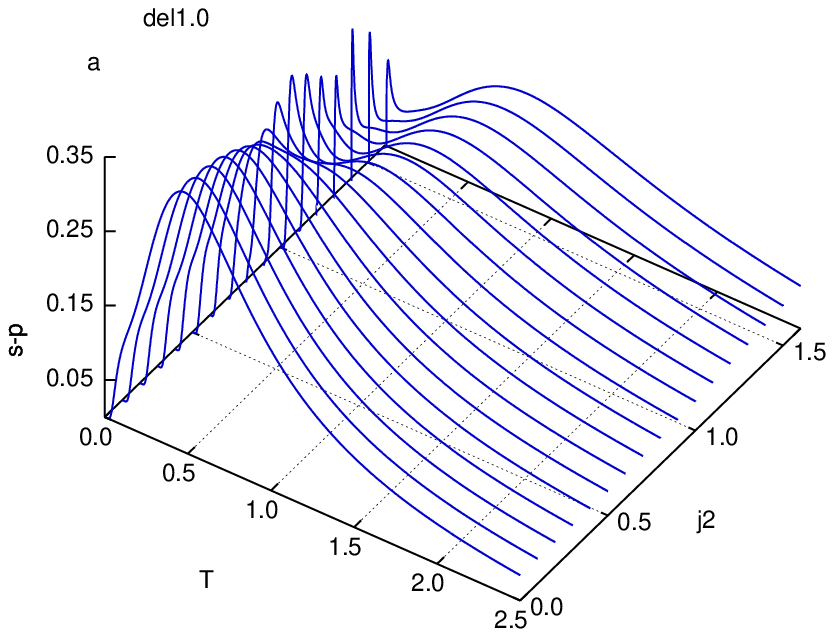}
\vskip -.5 in
\includegraphics[scale=0.80]{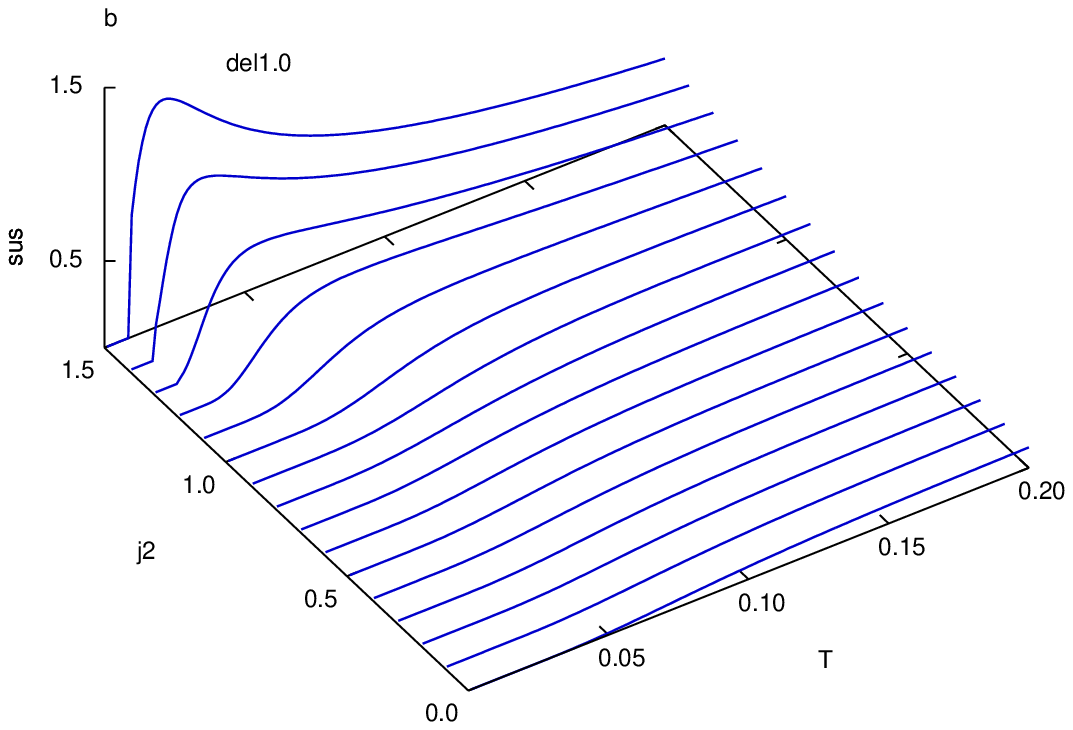}
\caption{\label{thermo}{(Color Online) 
Variation of specific heat, $C_{\rm V}/J_1$, (a) 
and susceptibility, $\chi/J_1$, (b)  with $k_{\rm B}T/J_1$ and 
$J_2/J_1$ for $\Delta = 1$.}}
\end{center}
\end{figure}

\begin{figure}[h]
\begin{center}
\psfrag{T}{\text{$k_{\rm B}T/J_1$}}
\psfrag{susc}{\text{$\chi/J_1$}}
\psfrag{a}{Kink-Antikink}
\psfrag{e}{Exact diagonalization}
\includegraphics[scale=0.70]{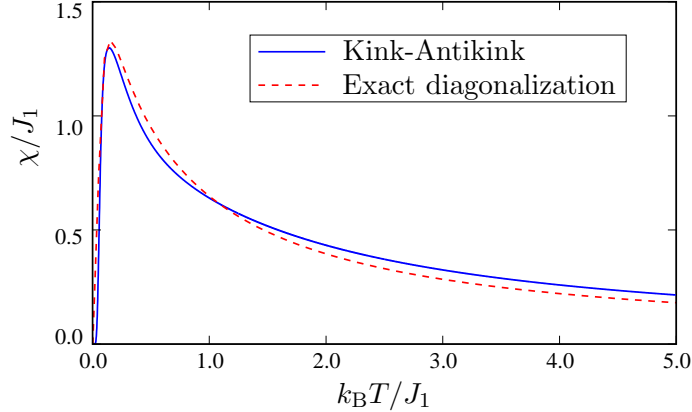}
\caption{\label{susc-comparison}{(Color Online) 
 Variation of specific heat, $C_{\rm V}/J_1$, with $k_{\rm B}T/J_1$ for 
$J_2/J_1=1$ and $\Delta = 1$. Exact diagonalization data is 
for the periodic chain of 20 ($N=10$) sites.}}
\end{center}
\end{figure}

Finally, specific heat, $C_{\rm V}$, and susceptibility, $\chi$, 
have been estimated numerically for $\Delta = 1$ and plotted with respect to $k_{\rm B}T/J_1$ 
and $J_2/J_1$, in Fig. \ref{thermo} (a) and (b), respectively. 
The appearance of additional peak 
in $C_{\rm V}$ at low temperatures with the increase of NNN bond strength, 
$J_2$ indicates the presence of frustration in the system. Position of the sharp 
peak moves toward low temperature with further increase of $J_2/J_1$ above $J_2/J_1=1$. 
This observation indicates the lowering of spin-gap in this region. 
On the other hand, a single broad peak 
appears in $\chi$, where height of the peak increases with the increase of $J_2/J_1$. 
In case of periodic chain, states with solitary kink or antikink cluster do not appear. 
Number of kink and antikink clusters are the same in the periodic chain and they appear 
alternately. Hence, in the low-energy excitations of the periodic chain, 
an antikink cluster must appear along with 
a kink cluster. For $J_2/J_1=1$ and $\Delta=1$, Kubo had indicated 
the existence of a 
non-dispersive state in the periodic sawtooth chain which one was identified 
as the kink state by Sen {\em et. al.} \cite{Kubo,Sen-Shastry}. 
In this case, energy of the non-dispersive kink coincides with the minimum of the 
antikink dispersion, which is $E_{\rm Gap}=0.2192J_1$. 
Based on this observation, expression of `Kink-Antikink' susceptibility 
has been derived in the article \cite{Sen-Shastry}. 
Following the same approach and using the antikink dispersion, $\omega_m(k)$, 
`Kink-Antikink' susceptibility has been obtained. This value of $\chi$ is compared with 
the numerical results estimated for the periodic chain (Fig. \ref{susc-comparison}). 
The numerical value of $\chi$ is five-time lower than the analytic value of that. 
However, the positions of the peaks for $\chi$ of those two estimations are the same, which 
is shown in Fig. \ref{susc-comparison}, where the numerical data is multiplied by five for comparison. 

\section{Conclusions}
\label{conclusion}
In this investigation, ground state energy, antikink dispersions, spin-gap, 
specific heat and susceptibility of an asymmetric $J_1$-$J_2$ sawtooth spin-$\frac{1}{2}$ 
anisotropic Heisenberg antiferromagnetic chain have been derived by using analytic 
and numerical methods in the full 
anisotropic regime $0\leq \Delta \leq 1$.
All of the above quantities have been obtained before for this model when $\Delta=1$ 
and $J_1=J_2$ \cite{Sen-Shastry,Nakamura} along with the spin-gap when $J_1\neq J_2$ \cite{Blundell}. 
However, in this study, the effect of both asymmetry and anisotropy of this system 
has been addressed. 
The exact value of ground state energy has been derived when $J_1=J_2$ for the entire 
anisotropic region $0\leq \Delta \leq 1$. 
Dispersions of three different antikinks spread over 1, 5 and 9 sites are evaluated. 
Spin-gap is estimated variationally out of these three antikinks. 
In addition to the spin-gap, specific heat and susceptibility are estimated numerically 
by using exact diagonalization. Analytic dispersions are obtained for the open chain, 
however, periodic boundary conditions are adopted in case of exact diagonalization 
to accommodate more spins in the chains. Differences of the values of spin-gap in those 
two approaches arise due to the missing of contribution from the kink-antikink pair states in the 
analytic approach. Like the case of antikinks as explained before, 
3- and 7-cluster kinks decompose into the 1- and 5-cluster kinks, respectively. 
Dispersions of 5- and 9-cluster kinks for the open chains 
when $J_1\neq J_2$ are not derived in this study. 
Excitations of the higher cluster kinks for the open chain 
and energies of kink-antikink bound states for the periodic chain 
will be addressed in the future study. 

This model behaves like a spin-1/2  
AFM Heisenberg chain with 2$N$ and $N$ sites, when $J_2/J_1=0$ 
and $J_2/J_1 \rightarrow \infty$, respectively. A non-zero spin-gap region 
thus persists in the intermediate zone bounded by these two 
extreme limits. In this investigation, region with non-zero 
spin-gap phase has been identified by using both analytic and numerical techniques. 
This phase does not attribute to the Haldane phase \cite{Haldane} since no 
string-order \cite{Rommelse} is found to develop 
in this system through the numerical studies. 
Existence of Haldane phase for the 
AFM spin-1/2 system has been predicted before in the bond-alternating 
Heisenberg chain \cite{Hida,Ghosh2}.
However, no long-range order of 
any kind of quantum correlations is in fact found to exist in this case.  

Behaviour of the system in the two extreme limits, $J_2/J_1=0$ 
and $J_2/J_1 \rightarrow \infty$, is different 
because of the presence of $N$ weakly interacting spins in the latter case. Although the 
spin-gap vanishes in both limits, the characteristic feature of 
specific heat and susceptibility in 
the limit, $J_2/J_1 \gg 1$, is not similar to that of the other limit,  
$J_2/J_1=0$, about the point $J_2/J_1=1$. 
Those differences attribute to the combined effect of frustration and presence of 
weakly interacting spins within the system. 

\acknowledgments 
AKG acknowledges fruitful discussion with Arghya Sil and 
financial support through the BRNS-sanctioned 
research project, 37(3)/14/16/2015, India. 
\appendix
\section{Matrix elements of $H$}
 \label{A1}
\bea
{}_1\langle 2n|H|2m\rangle_1&=&\frac{1}{2}g(\Delta)\big[J_1\left(\delta_{n,m}
-N(-1/2)^{|n-m|}\right)\nonumber \\
&&+(J_1\!-\!J_2)\big((1-|n-m|)(-1/2)^{|n-m|}-\delta_{n,m}\big)\big],\nonumber \\
{}_2\langle 2n|H|2m\rangle_2&=&g(\Delta)\big[J_1\big(\delta_{n,m}-(\delta_{n,m+2}+\delta_{n,m-2})/8 +N \{-3/4\, \delta_{n,m}
\!+\!3/16(\delta_{n,m+1}+\delta_{n,m-1}) \nonumber \\
&&+(-1/2)^{|n-m|+2}\} \big) +(J_1\!-\!J_2)\big(2\,\delta_{n,m}-3/4(\delta_{n,m+1}\!+\!\delta_{n,m-1})\nonumber \\
&&+3/16(\delta_{n,m+2}\!+\!\delta_{n,m-2})\!+\!(|n-m|\!-\!6)(-1/2)^{|n-m|+2} \big) \big],\nonumber \\
{}_4\langle 2n|H|2m\rangle_4&=&g(\Delta)\big[J_1\big(3/2\delta_{n,m}
-(\delta_{n,m+2}+\delta_{n,m-2})/8+(\delta_{n,m+4}+\delta_{n,m-4})/32\nonumber \\
&&+N \{-3/8\, \delta_{n,m}\!-\!3/64(\delta_{n,m+1}+\delta_{n,m-1})
\!+\!3/32(\delta_{n,m+2}+\delta_{n,m-2}) \nonumber \\
&&\!-\!3/64(\delta_{n,m+3}+\delta_{n,m-3})+(-1/2)^{|n-m|+3}\} \big)  +(J_1\!-\!J_2)\big(-7/8\,\delta_{n,m} \nonumber \\
&&+9/16(\delta_{n,m+1}\!+\!\delta_{n,m-1})-15/32(\delta_{n,m+2}+\delta_{n,m-2})+(\delta_{n,m+3}+\delta_{n,m-3})/4\nonumber \\
&&-(\delta_{n,m+4}+\delta_{n,m-4})/32\!+\!(|n-m|\!-\!11)(-1/2)^{|n-m|+3} \big) \big]. \nonumber
\eea
$g(\Delta)=1+\Delta/2$, and ${}_3\langle 2n|H|2m\rangle_3={}_2\langle 2n|H|2m\rangle_2$. 

When $N\rightarrow \infty$, 
\bea
{}_1\langle k|H|k\rangle_1&=&E_{\rm G}\;{}_1\langle k|k\rangle_1 + \frac{1}{2}g(\Delta)\big[J_1-(J_1\!-\!J_2)h(\cos{k})\big],\nonumber \\
{}_2\langle k|H|k\rangle_2&=&E_{\rm G}\;{}_2\langle k|k\rangle_2 + g(\Delta)\bigg[J_1\left(1-\frac{1}{4}\cos{2k}\right)
+(J_1\!-\!J_2)\bigg(2-\frac{3}{2}\cos{k}\nonumber \\
&&+\frac{1}{4}\cos{2k}-\frac{3}{2}f(\cos{k})+\frac{1}{4}h(\cos{k})\bigg)\bigg], \nonumber \\
{}_4\langle k|H|k\rangle_4&=&E_{\rm G}\;{}_4\langle k|k\rangle_4 + g(\Delta)\bigg[J_1\bigg(\frac{3}{2}-\frac{1}{4}\cos{2k}
+\frac{1}{16}\cos{4k}\bigg)+(J_1\!-\!J_2)\bigg(-\frac{7}{8}\nonumber \\
&&+\frac{9}{8}\cos{k}-\frac{15}{16}\cos{2k}+\frac{1}{2}\cos{3k}-\frac{1}{8}\cos{4k}
-\frac{11}{8}f(\cos{k})-\frac{1}{8}h(\cos{k})\bigg)\bigg]. \nonumber 
\eea
$h(x)=4(4+5x)/(25+40x+16x^2)$. 
\section{Additional matrix elements for obtaining variational dispersion relation}
 \label{A2}
\bea
{}_1\langle 2n|2m\rangle_2 &+&{}_1\langle 2n|2m\rangle_3=(-1/2)^{|n-m|+1},\nonumber \\
{}_1\langle 2n|2m\rangle_4 &=&3/4\, \delta_{n,m}
-3/8(\delta_{n,m+1}\!+\!\delta_{n,m-1})+(-1/2)^{|n-m|+1},\nonumber \\
{}_2\langle 2n|2m\rangle_3 &+&{}_3\langle 2n|2m\rangle_2=-9/4\, \delta_{n,m}
+3/4(\delta_{n,m+1}\!+\!\delta_{n,m-1})+5(-1/2)^{|n-m|+2},\nonumber \\
{}_2\langle 2n|2m\rangle_4 &+&{}_3\langle 2n|2m\rangle_4=-3/8\, \delta_{n,m}
+3/8(\delta_{n,m+1}\!+\!\delta_{n,m-1})+(-1/2)^{|n-m|+2},\nonumber\\
{}_1\langle 2n|H|2m\rangle_2 &+&{}_1\langle 2n|H|2m\rangle_3=g(\Delta)\big[NJ_1(-1/2)^{|n-m|+2}
+(J_1\!-\!J_2)\big(3/4\, \delta_{n,m}\nonumber\\
&&+(|n\!-\!m|\!-\!2)(-1/2)^{|n-m|+2} \big)\big],\nonumber\\
{}_1\langle 2n|H|2m\rangle_4 &=&g(\Delta)\big[J_1\big(1/8\, \delta_{n,m}-1/8(\delta_{n,m+2}\!+\!\delta_{n,m-2})
+N\{-3/8 \, \delta_{n,m}+3/16\,(\delta_{n,m+1}\!+\!\delta_{n,m-1}) \nonumber\\
&&+(-1/2)^{|n-m|+2}\} \big) +(J_1\!-\!J_2)\big(7/4\,\delta_{n,m}-3/4(\delta_{n,m+1}\!+\!\delta_{n,m-1})\nonumber \\
&&+1/8(\delta_{n,m+2}\!+\!\delta_{n,m-2})\!+\!(|n\!-\!m|\!-\!6)(-1/2)^{|n-m|+2} \big) \big],\nonumber \\
{}_2\langle 2n|H|2m\rangle_3 &+&{}_3\langle 2n|H|2m\rangle_2=g(\Delta)\big[J_1\big(-\delta_{n,m}+1/4(\delta_{n,m+2}\!+\!\delta_{n,m-2})
+N\{7/4\, \delta_{n,m}\nonumber\\&&-3/8\,(\delta_{n,m+1}\!+\!\delta_{n,m-1}) 
+5(-1/2)^{|n-m|+3}\} \big) +(J_1\!-\!J_2)\big(-29/4\,\delta_{n,m}\nonumber \\
&&+3/2(\delta_{n,m+1}\!+\!\delta_{n,m-1})-1/4(\delta_{n,m+2}\!+\!\delta_{n,m-2})\!+\!(5|n\!-\!m|\!-\!27)(-1/2)^{|n-m|+3} \big) \big],\nonumber \\
{}_2\langle 2n|H|2m\rangle_4 &+&{}_3\langle 2n|H|2m\rangle_4=g(\Delta)\big[J_1\big(1/2\,(\delta_{n,m+1}\!+\!\delta_{n,m-1})-1/8\,
(\delta_{n,m+3}\!+\!\delta_{n,m-3})\nonumber\\
&&+N\{9/16\, \delta_{n,m}-3/8\,(\delta_{n,m+1}\!+\!\delta_{n,m-1}) +3/32\,(\delta_{n,m+2}\!+\!\delta_{n,m-2})
+(-1/2)^{|n-m|+2}\} \big)\nonumber \\&& +(J_1\!-\!J_2)\big(9/4\,\delta_{n,m}
-3/4(\delta_{n,m+1}\!+\!\delta_{n,m-1})
+13/32\,(\delta_{n,m+2}\!+\!\delta_{n,m-2})\nonumber \\
&&-1/16\,(\delta_{n,m+3}\!+\!\delta_{n,m-3})\!-\!(|n\!-\!m|\!-\!9)(-1/2)^{|n-m|+3} \big) \big].\nonumber 
\eea
Therefore, when $N\rightarrow \infty$
\bea
{}_1\langle k|k\rangle_2 &\!+\!&{}_1\langle k|k\rangle_3=-\frac{1}{2}f(\cos{k}),\;
{}_1\langle k|k\rangle_4=\frac{3}{4}(1-\cos{k})-\frac{1}{2}f(\cos{k}),\nonumber \\
{}_2\langle k|k\rangle_3 &\!+\!&{}_3\langle k|k\rangle_2=-\frac{9}{4}+\frac{3}{2}\cos{k}+\frac{5}{4}f(\cos{k}),\;
{}_2\langle k|k\rangle_4 \!+\!{}_3\langle k|k\rangle_4=-\frac{3}{8}+\frac{3}{4}\cos{k}+\frac{1}{4}f(\cos{k}),\nonumber \\
{}_1\langle k|H|k\rangle_2 &\!+\!&{}_1\langle k|H|k\rangle_3=E_{\rm G}({}_1\langle k|k\rangle_2 \!+\!{}_1\langle k|k\rangle_3)
+g(\Delta)(J_1\!-\!J_2)\bigg(\frac{1}{4}-\frac{1}{2}f(\cos{k})+\frac{1}{4}h(\cos{k})\bigg),\nonumber \\
{}_1\langle k|H|k\rangle_4&=&E_{\rm G}\,{}_1\langle k|k\rangle_4 + g(\Delta)\bigg[\frac{J_1}{4}\bigg(\frac{1}{2}-\cos{2k}\bigg) 
\!+\!(J_1\!-\!J_2)\bigg(\frac{7}{4}\!-\!\frac{3}{2}\cos{k}\!+\!\frac{1}{4}\cos{2k}\nonumber \\
&&-\frac{3}{2}f(\cos{k})\!+\!\frac{1}{4}h(\cos{k})  \bigg)\bigg],\nonumber \\
{}_2\langle k|H|k\rangle_3 &+&{}_3\langle k|H|k\rangle_2=E_{\rm G}({}_2\langle k|k\rangle_3 \!+\!{}_3\langle k|k\rangle_2)
+g(\Delta)\bigg[\frac{J_1}{2}\bigg(-2+\cos{2k}\bigg)\nonumber \\
&&+(J_1\!-\!J_2)\bigg(-\frac{31}{8}+3\cos{k}-\frac{1}{2}\cos{2k}-\frac{27}{8}f(\cos{k})
+\frac{5}{8}h(\cos{k})\bigg)\bigg],\nonumber \\
{}_2\langle k|H|k\rangle_4 &+&{}_3\langle k|H|k\rangle_4=E_{\rm G}({}_2\langle k|k\rangle_4 \!+\!{}_3\langle k|k\rangle_4)
+g(\Delta)\bigg[\frac{J_1}{2}\cos{k}\nonumber \\
&&+(J_1\!-\!J_2)\bigg(-\frac{17}{16}+\frac{5}{4}\cos{k}-\frac{1}{8}\cos{2k}+\frac{7}{8}f(\cos{k})
-\frac{1}{8}h(\cos{k})\bigg)\bigg],\nonumber 
\eea

\end{document}